\documentclass[12pt,letter]{amsart}
\usepackage{xcolor}
\usepackage[all,color]{xy}
\usepackage{subcaption,tensor,appendix,import,stmaryrd,wasysym,amsmath,amssymb,amsthm,amscd,amsxtra,amsfonts,mathrsfs,graphicx,enumerate,bm,slashed,array,setspace,enumitem,minibox,booktabs,rotating,multirow,adjustbox,hhline,arydshln,placeins}
\input xy
\xyoption{all}
\newtheorem{Theorem}{Theorem}[section]
\newtheorem{Lemma}[Theorem]{Lemma}
\newtheorem{Proposition}[Theorem]{Proposition}
\newtheorem{Corollary}[Theorem]{Corollary}
\newtheorem{Example}[Theorem]{Example}
\newtheorem{Remark}[Theorem]{Remark}

\newtheorem{Definition}[Theorem]{Definition}

\newtheorem{Notation}[Theorem]{Notation}
\newtheorem*{Definition*}{Definition}
\DeclareMathOperator{\im}{im}
\DeclareMathOperator{\vol}{vol}

\advance\evensidemargin-.5in
\advance\oddsidemargin-.5in
\advance\textwidth1in

\definecolor{darkgreen}{rgb}{0.0, 0.5, 0.0}
\definecolor{darkred}{rgb}{0.7, 0.11, 0.11}

\begin{document}

\author{Charlie Beil}
\address{Institut f\"ur Mathematik und Wissenschaftliches Rechnen, Universit\"at Graz, Heinrichstrasse 36, 8010 Graz, Austria.}
 \email{charles.beil@uni-graz.at}
\title[Spacetime geometry of spin, polarization, and wavefunction collapse]{Spacetime geometry of spin, polarization,\\ and wavefunction collapse}

\begin{abstract}
To incorporate quantum nonlocality into general relativity, we propose that the preparation and measurement of a quantum system are simultaneous events. 
To make progress in realizing this proposal, we introduce a spacetime geometry that is endowed with particles which have no distinct points in their worldlines; we call these particles `pointons'.
This new geometry recently arose in nonnoetherian algebraic geometry. 
We show that on such a spacetime, metrics are degenerate and tangent spaces have variable dimension. 
This variability then implies that pointons are spin-$\tfrac 12$ fermions that satisfy the Born rule, where a projective measurement of spin corresponds to an actual projection of tangent spaces of different dimensions. 
Furthermore, the $4$-velocities of pointons are necessarily replaced by their Hodge duals, and this transfer from vector to pseudo-tensor introduces a free choice of orientation that we identify with electric charge. 
Finally, a simple composite model of electrons and photons results from the metric degeneracy, and from this we obtain a new ontological model of photon polarization. 
\end{abstract}

\maketitle

\tableofcontents

\section{Simultaneity of preparation and measurement}

In this article we introduce the idea that the fundamental source of quantum phenomena is that clocks in the same inertial frame may tick at different rates. 
Specifically, we propose that time does not advance in a quantum system undergoing unitary evolution.
Consequently, \textit{the preparation and measurement of a quantum system are simultaneous events.}
This simultaneity is then the origin of quantum nonlocality. 
Our aim here is to make this idea precise, both mathematically and physically, by modifying the geometry of classical general relativity. 

The spacetime model we will introduce produces many of the standard features of quantum theory, but does not reproduce quantum theory exactly. 
In particular, we model quantum phenomena not with Hilbert spaces, but with \textit{degenerate} spacetime metrics obtained from our new notion of simultaneity. 
We will show that such metrics yield a composite model of electrons and photons, as well as ($\psi$-epistemic ontological models of) spin, polarization, and their wavefunction collapse.
In companion articles we reproduce photon path superposition and entanglement using this framework \cite{B7}, and extend our composite model of electrons and photons introduced here to all standard model particles \cite{B1,B2}. 

We remark that the original motivation for the development of this new geometry, which we call `nonnoetherian geometry', arose from brane tilings in string theory (e.g., \cite{B6, F-W}).\footnote{A `brane tiling' (or `dimer model') is a directed graph that embeds in a torus with certain conditions.
Associated to a brane tiling is a quiver gauge theory whose vacuum moduli space is either a toric algebraic variety (in which case the brane tiling is said to be `consistent' and the gauge theory is superconformal; see e.g., \cite{Br,D,DHP}) or a nonnoetherian scheme \cite{B3} (in which case the brane tiling is `inconsistent' and the gauge theory is not superconformal).
The former gauge theories are stable, the latter are unstable, and both are physically allowed; in M-theory the two kinds of theories are on an equal footing.
It had always been assumed, however, that nonnoetherian schemes do not admit concrete geometric descriptions---something that can be visualized---but rather are only abstract constructions. 
With the aim of a geometric picture for the nonnoetherian moduli spaces, the author introduced nonnoetherian algebraic geometry as the geometry of algebraic varieties with positive dimensional `smeared-out' points \cite{B4,B5}.
Under this new framework, the moduli space of an inconsistent brane tiling admits an honest geometric description as a toric algebraic variety with an embedded curve or surface that is identified as a single closed point \cite{B6}.} 
However, here we apply nonnoetherian geometry directly to general relativity.\\
\\
\textbf{Notation:} Tensors labeled with upper and lower indices $a,b, \ldots$ represent covector and vector slots respectively in Penrose's abstract index notation (so $v^a \in V$ and $v_a \in V^*$), and tensors labeled with indices $\mu, \nu, \ldots$ denote components with respect to a coordinate basis.
We denote the tangent space of a manifold $M$ at $p \in M$ by $M_p := T_pM$ and its (vector space) dimension by $\dim M_p$.
Given a curve $\beta: I \to M$, we will often denote its image $\beta(I)$ also by $\beta$. 
We use natural units, $c = 1$, and the signature $(+,-,-,-)$ throughout.

\section{Spacetime with a degenerate metric}

We would like to modify general relativity so that the preparation and measurement of a quantum system are simultaneous events.
To this end, we introduce the following.

\begin{Definition} \label{nstdef} \rm{
Let $(\tilde{M},g)$ be an orientable Lorentzian $4$-manifold.
Consider a locally finite set of particles on $\tilde{M}$ with worldlines $\beta_i \subset \tilde{M}$.
We call the set 
\begin{equation*}
M := (\tilde{M} \setminus (\cup_i \beta_i)) \cup (\cup_j \{\beta_j \}),
\end{equation*}
where each $\beta_j$ is a single point of $M$, 
an \textit{internal spacetime}, or simply \textit{spacetime}. 
We call $\tilde{M}$ the \textit{external spacetime} of $M$, and the particles \textit{pointons}.
}\end{Definition}

We thus construct spacetime $M$ from $\tilde{M}$ by replacing each $\beta_i$ in $\tilde{M}$ with a single point.
Although points are usually defined to be $0$-dimensional objects (an exception being the points of a scheme in algebraic geometry), 
this is not the case in our construction of $M$.
We call a pointon worldline $\beta$ a `point' because there are no ($0$-dimensional) points properly contained inside of it in $M$.
However, we continue to regard $\beta$ as a curve, that is, as a $1$-dimensional object, since it appears to be a curve in a deleted neighborhood about it.
In other words, in constructing $M$ we do not contract $\beta \subset \tilde{M}$ to a $0$-dimensional point, as is done when taking geometric or topological quotients. 
Instead, we keep the embedding of $\beta$ intact by using the metric on $\tilde{M}$ to measure distances near $\beta$ on $M$ (made precise below).
Informally, we do not `throw away' $\tilde{M}$ in forming $M$.
The worldline $\beta$ is then a continuum of distinct $0$-dimensional points in $\tilde{M}$, and a single $1$-dimensional point in $M$.
In particular, \textit{time does not advance along $\beta$ in spacetime $M$.}

Since a pointon worldline $\beta$ is a single point of spacetime $M$, \textit{we cannot define a tangent vector $v \in \tilde{M}_{\beta(t)}$ (or $4$-velocity) along $\beta$ in $M$.}
Furthermore, any `metric' $h$ on $M$, however it may be defined, should be equal on any vectors $w_1, w_2 \in \tilde{M}_p$ that map to points $\exp_p (w_1), \exp_p (w_2) \in \tilde{M}$ which are identified in $M$ and lie in a sufficiently small open neighborhood of $p$ in $\tilde{M}$.

\begin{Lemma} \label{equiv 0}
Let $\beta \subset \tilde{M}$ be a pointon worldline with tangent vector $v$.
For $p \in \tilde{M}$, let $h_{ab}: \tilde{M}_p \otimes \tilde{M}_p \to \mathbb{R}$ be a symmetric tensor such that $h_{ab}w_1^a = h_{ab}w_2^a$ whenever the points $\exp_p (w_1), \exp_p (w_2) \in \tilde{M}$ are identified in $M$ but are not equal in $\tilde{M}$.
Then $h_{ab}v^a \equiv 0$ along any geodesic segment of $\beta$. 
\end{Lemma}

\begin{proof}
Without loss of generality, we may suppose $\beta(I)$ is a geodesic segment of $\beta$ with $t = 0$ in the interval $I \subset \mathbb{R}$. 
Then the points $\exp_{\beta(0)} (\lambda v)$ and $\exp_{\beta(0)}(\lambda' v)$ lie in $\beta(I) \subset \tilde{M}$ for all $\lambda, \lambda' \in I$, and are thus identified in $M$. 
Whence, for all $w \in \tilde{M}_{\beta(0)}$,
\begin{equation*}
\lambda h(v,w) = h(\lambda v, w) = h(\lambda' v, w) = \lambda' h(v,w).
\end{equation*}
Therefore, by choosing $\lambda \not = \lambda'$ we find $h_{ab}v^a = h(v,-) = 0$.
\end{proof}

Lemma \ref{equiv 0} implies that to construct an internal metric $h_{ab}$ at $p \in \tilde{M}$ from the metric $g_{ab}$ on $\tilde{M}$, it must project out each vector $v$ tangent to a geodesic pointon worldline $\beta \subset \tilde{M}$ at $p$.
Let $v$ be such a vector. 
If $v$ is a timelike resp.\ spacelike unit vector, then the orthogonal projection of $v$ is
\begin{equation} \label{projection of v}
[v]_{ab} := g_{ab} - v_av_b \ \ \ \ \text{ resp.\ } \ \ \ \ [v]_{ab} := g_{ab} + v_av_b.
\end{equation}
However, if $v$ is null, then (\ref{projection of v}) does not project out $v$.

\begin{Lemma} \label{null projection lemma}
Let $v$ be a null vector.
Any minimal orthogonal projection $\tensor{[v]}{^a_b}$ of $v$ is of the form
\begin{equation} \label{null projection}
[v]_{ab} := g_{ab} + v_av'_b + v_bv'_a,
\end{equation}
where $v'$ is a null vector satisfying $v^a v'_a = -1$.
\end{Lemma}

\begin{proof}
It suffices to suppose
\begin{equation*}
\tensor{[v]}{^a_b} = \tensor{\delta}{^a_b} - c v^a v'_b + \tensor{A}{^a_b}
\end{equation*}
for some $c \in \mathbb{R}$, vector $v'$, and tensor $\tensor{A}{^a_b}$.
Since $\tensor{[v]}{^a_b}$ is a projection, its indices are symmetric, $[v]_{ab} = [v]_{ba}$, and so $\tensor{A}{^a_b} = -c v'^av_b + \cdots$.
Moreover, since the rank of $\tensor{[v]}{^a_b}$ is maximum, we have $\tensor{A}{^a_b} = -c v'^av_b$.
Whence,
\begin{equation*}
\tensor{[v]}{^a_b} = \tensor{\delta}{^a_b} - c v^a v'_b - c v'^av_b.
\end{equation*}
Since $v$ is null and $\tensor{[v]}{^a_b}v^b = 0$, $v$ and $v'$ are linearly independent.
Without loss of generality we may choose $v^av'_a = -1$.
Then $c = -1$, again since $\tensor{[v]}{^a_b}v^b = 0$.

Now assume to the contrary that $v'$ is not null.
Then $\tensor{[v]}{^a_b} v'^b = (v'^2) v^a \not = 0$ and $\tensor{[v]}{^a_b}\tensor{[v]}{^b_c} v'^c = (v'^2) \tensor{[v]}{^a_b} v^b = 0$. 
But this implies $[v]^2 \not = [v]$, a contradiction since $[v]$ is an orthogonal projection.
Therefore $v'$ must also be null.
\end{proof}

We will determine the physical meaning of the null vector $v'$ in Section \ref{photons section}.

\begin{Definition} \label{internal metric def} \rm{
Fix $p \in \tilde{M}$.
We call the Lorentzian metric $g_{ab}: \tilde{M}_p \otimes \tilde{M}_p \to \mathbb{R}$ an \textit{external metric} at $p$.\footnote{The external metric $g_{ab}$ is used to identify the tangent and cotangent spaces $\tilde{M}_p$ and $\tilde{M}_p^*$, and thus to raise and lower indices.} 
Let $v_1, \ldots, v_n \in \tilde{M}_p$ be the tangent vectors to the pointon worldlines $\beta_1, \ldots, \beta_n$ at $p$.
We define the corresponding \textit{internal metric} to be the degenerate symmetric rank-$2$ tensor given by the composition of projections (\ref{projection of v}) and (\ref{null projection}),
\begin{equation*} \label{projection}
h = h_p = \tensor{h}{^a_b} := \tensor{[v_1]}{^a_c}\tensor{[v_2]}{^c_d} \cdots \tensor{[v_n]}{^e_b}: \tilde{M}^*_p \otimes \tilde{M}_p \to \mathbb{R}.
\end{equation*}
}\end{Definition}

As is well known, for any Riemannian manifold $\tilde{M}$, there is a geodesic ball $B_p$ about each point $p \in \tilde{M}$ for which the exponential map $\operatorname{exp}_p: \tilde{M}_p \to \tilde{M}$ is injective.
By our definition of internal metric, this property continues to hold on internal spacetimes.

\begin{Definition} \label{internal tangent space def} \rm{
The \textit{internal tangent space} at a point $p \in \tilde{M}$ is the image of $h$ at $p$,
\begin{equation*}
M_p := \im h = \im (\tensor{h}{^a_b}) = \{ v^a \in \tilde{M}_p \, | \, \tensor{h}{^a_b}\tensor{v}{^b} = \tensor{v}{^a} \} \subseteq \tilde{M}_p,
\end{equation*}
and thus is a subspace of the $4$-dimensional tangent space $\tilde{M}_p$.
We call a vector $v \in \tilde{M}_p$ \textit{internal} if $v$ is in $M_p$, and \textit{external} otherwise. 
}\end{Definition}

We will find in the next sections that \textit{the variability of the dimensions of the internal tangent spaces $M_p$ is the source of spin, polarization, and their wavefunction collapse.}

\section{Internal $4$-velocities: electric charge and spin}

We first recall orientation and Hodge duality from Maxwell's equations to establish notation (e.g., \cite{F, N}).

An orientation of a vector space $V$ is given by fixing an ordered basis $\mathcal{B}$ of $V$, and declaring any ordered basis to be positive (resp.\ negative) if it can be obtained from $\mathcal{B}$ by a base change with a positive (resp.\ negative) determinant.

Let $M$ be an internal spacetime with external spacetime $\tilde{M}$ as in Definition \ref{nstdef}; in particular, $\tilde{M}$ is a $4$-dimensional Lorentzian manifold. 
Fix $p \in \tilde{M}$ and an (arbitrary) orientation for the tangent space $\tilde{M}_p$.
Let $e_0, \ldots, e_3 \in \tilde{M}_p$ be a positive orthonormal tetrad.
The exterior algebra $\bigwedge \tilde{M}_p^* := \oplus_{n = 0}^4 \bigwedge^n \tilde{M}_p^*$ has basis consisting of $1$ and the set of volume forms
\begin{equation*} \label{volume form}
\vol ( \operatorname{span} \{ e_{j_1}, \ldots, e_{j_n} \}) := e^{j_1} \wedge \cdots \wedge e^{j_n} \in {\bigwedge} \! ^n \, \tilde{M}_p^*
\end{equation*}
of the subspaces $\operatorname{span} \{ e_{j_1}, \ldots, e_{j_{n}} \} \subseteq \tilde{M}_p$, with $1 \leq j_1 < \cdots < j_{n} \leq 4$.
In particular,\footnote{The metric appears in the volume form (\ref{volume form}) in arbitrary local coordinates $x^{\mu}$ on $\tilde{M}$ near $p$, 
$\vol (\tilde{M}_p) = o \left| \det e \right| dx^0 \wedge \cdots \wedge dx^3 = o \sqrt{ \left| \det g \right| } dx^0 \wedge \cdots \wedge dx^3$,
where $o \in \{ \pm 1 \}$ is the orientation of $\tilde{M}_p$.}
\begin{equation} \label{volume form}
\vol (\tilde{M}_p) = e^0 \wedge e^1 \wedge e^2 \wedge e^3 \in {\bigwedge} \! ^4 \, \tilde{M}_p^*.
\end{equation} 
The Hodge dual $\star \psi$ of an element $\psi \in \bigwedge \tilde{M}_p^*$ is defined on the basis elements $\psi = e^{j_1} \wedge \cdots \wedge e^{j_{n}}$ by 
\begin{equation*} \label{Hodge dual def}
\psi \wedge \star \psi = \det  [ \eta_{j_i j_k} ] \vol (\tilde{M}_p),
\end{equation*}
and extended $\mathbb{R}$-multilinearly to $\bigwedge \tilde{M}_p^*$.
For example, $\star 1 = \vol (\tilde{M}_p)$.
Note that Hodge duals are \textit{pseudo}-forms, that is, they depend on a choice of orientation, since they are defined using the volume form $\vol (\tilde{M}_p)$.
This fact will play a fundamental role in our framework.

Now consider a pointon with timelike worldline $\beta \subset \tilde{M}$ and $4$-velocity $v$ on $\tilde{M}$. 
Since time does not advance along $\beta$, $v$ is orthogonal to spacetime $M$: $h(v) = 0$. 
We would therefore like to replace $v$ with a new geometric object $\breve{v}$ that is intrinsic to spacetime $M$ and independent of $\tilde{M}$.

Naively, we may replace $v \in \ker h \subset \tilde{M}_{\beta(t)}$ at a point $\beta(t) \in \tilde{M}$ with the Hodge dual $\star \! \vol (\ker h)$ of the volume form of the kernel $\ker h \subset \tilde{M}_{\beta(t)}$. 
Indeed, each $1$-form in $\star \! \vol (\ker h)$ lies in the image $\im h = M_{\beta(t)}$, since $\im h$ is orthogonal to $\ker h$ in $\tilde{M}_p$.
However, a Hodge dual is a \textit{pseudo}-form, and thus depends on a choice of orientation of the subspace $\ker h$. 

To make our geometric replacement $\breve{v}$ of $v$ fully intrinsic to $M$, we need to eliminate this dependency by allowing an orientation of $\ker h$ to be freely chosen, independent of any (non-physical) choice of orientation of $\tilde{M}_{\beta(t)}$.
Thus, by requiring $\breve{v}$ to be intrinsic to $M$ we obtain a new $\mathbb{Z}_2$ parameter.
Fundamentally, this parameter arises because we are replacing a vector, which does not depend on any choice of orientation, with a pseudo-tensor that does.

\begin{Definition} \label{internal 4-velocity def} \rm{
We define the \textit{internal $4$-velocity} of a pointon at $p = \beta(t) \in \tilde{M}$ with $4$-velocity $v \in \ker h \subset \tilde{M}_p$ to be the pseudo-form
\begin{equation*} \label{internal form}
\breve{v}_{a \cdots b} := o_{\operatorname{ker}h} \star \! \vol (\ker h) \in {\bigwedge} \! ^{\dim M_p} \, M_p^*,
\end{equation*}
where $o_{\operatorname{ker}h} \in \{ \pm 1 \}$ is a free parameter independent of any orientation of $\tilde{M}_p$.
Note that the rank of $\breve{v}$ changes along $\beta \subset \tilde{M}$ whenever the dimension of $M_{\beta(t)}$ changes.
}\end{Definition}

Again consider a pointon with timelike worldline $\beta \subset \tilde{M}$ and $4$-velocity $v$.
Choose a positive orthonormal tetrad $e_0, \ldots, e_3 \in \tilde{M}_{\beta(t)}$ along $\beta$ for which $e_0 = v$ and fix a point $p = \beta(t) \in \tilde{M}$.

(i) $\dim M_p = 3$: First suppose that $\beta$ does not transversely intersect another pointon worldline at $p$.
Then the internal $4$-velocity at $p$ is 
\begin{equation*} 
\breve{v}^{abc} = o_0 \, e_1 \wedge e_2 \wedge e_3,
\end{equation*}
where $o_0 \in \{ \pm 1 \}$ is a free choice of time orientation (in the rest frame of the pointon), independent of any orientation of $\tilde{M}_p$.
We identify this timelike orientation with the electric charge of the pointon,
\begin{equation} \label{electric charge}
o_0 \in \{ \pm e \}.
\end{equation}
This is similar to the St\"uckelberg-Feynman interpretation of antimatter in quantum field theory, where positrons are regarded as electrons that travel backwards in time \cite{S}.
We note, however, that in contrast to the St\"uckelberg-Feynman interpretation, \textit{time is stationary along the pointon's worldline $\beta$: since $\beta$ is a single point of spacetime $M$, the pointon does not travel backwards along $\beta$ just as it does not travel forwards.}

(ii) $\dim M_p = 1$: Suppose $\beta$ transversely intersects other pointon worldlines at $p$ such that $\dim M_p = 1$.
If we choose a tetrad for which $\ker h = \operatorname{span}\{ e_0, e_1, e_2 \}$ at $p$, then the internal $4$-velocity is the vector
\begin{equation*}
\breve{v}^a = o_0 o_P \, e_3,
\end{equation*}
where $o_P \in \{ \pm 1 \}$ is a free choice of orientation of the plane $P = \operatorname{span}\{ e_1, e_2 \}$ orthogonal to $\breve{v}^a$ and independent of any orientation of $\tilde{M}_p$.
We call the internal $4$-velocity $\breve{v}^a$ the \textit{epistemic spin vector} of the pointon at $p$, and identify the choice of orientation $o_P$ as spin, up $\uparrow$ or down $\downarrow$, in the $e_3$ direction,
\begin{equation*}
o_P \in \{ \uparrow, \downarrow \}.
\end{equation*}
We then parallel transport $\breve{v}^a$ along $\beta$ until it is projected under $h$ onto a subsequent $1$-dimensional subspace $M_{\beta(t')}$.
In the next section we will show how this projection yields a realist model of spin wavefunction collapse. 

\begin{Remark} \label{composite standard model} \rm{
The identification of time orientation with electric charge yields a new physical interpretation of spinors, wherein spinor chirality becomes electric charge \cite[Proposition 2.2]{B2}.
This new interpretation in turns yields a derivation of the standard model particles from the free Dirac Lagrangian \cite[Theorem 7.1]{B2}.
In particular, it establishes a composite model (interpretation) of the standard model, where its particles are represented by bound states of pointons.
\textit{This model has only one fundamental particle, namely the pointon, and only one fundamental interaction, namely pair creation and annihilation of pointons of opposite charge.}
We note, however, that the compositeness may be regarded as purely mathematical (similar to a direct sum decomposition of a representation into irreducibles), since \textit{the pointons in a bound state share the same wordline}.
In any case, this feature makes our model quite different from other composite (or preon) models and eliminates the preon problem of finite particle size.

The free Dirac Lagrangian specifies both the possible pointon bound states and, with two additional `fusion rules', their possible interactions (the standard model trivalent vertices) \cite[Theorem B]{B1}, \cite[Section 5; Theorem 5.9]{B2}.\footnote{Although our model reproduces much of the standard model, a few differences remain.
Foremost, our model predicts one additional new particle: a massive neutral spin-$2$ boson \cite[Section 7]{B2}. 
The trivalent vertices that involve this boson are derived in \cite[Section 6]{B1}. 
It is possible, however, to impose extra `natural' conditions that would exclude this particle without affecting the standard model particles \cite[Remark 7.2]{B2}.
Moreover, our model predicts subtle differences in the allowed polarization/spin states that may occur in a few trivalent vertices; these differences are derived in \cite[Table 2; Section 5]{B1} and should make our model experimentally falsifiable.}
In this framework, then, gauge theory is regarded as an approximate description of the Dirac Lagrangian.
Furthermore, the neutral standard model particles (the photon, $Z$ boson, Higgs boson, and neutrinos) are represented by bound states of an even number of pointons, half with positive charge and half with negative charge.\footnote{We mention that a pointon without spin is derived from the model, but such a pointon can only exist in a bound state with another pointon possessing spin and opposite charge (a $\gamma^0$ eigenspinor \cite[Lemma 5.8]{B2}).
This state represents the electron neutrino and is equivalent to a spinor that is supported on a worldline which is not a pointon; see \cite[footnote 9]{B2}.}
In Section \ref{photons section} we will introduce this model for electrons, positrons, and photons.

Finally, the possible free choices of orientation that may arise from vanishing subspaces are given in Table \ref{orientation table}.
In this article we only consider electric charge and spin.
We note that the choice of orientation $o_{0123} \in \{ \pm 1\}$ for any external tangent space is nonphysical.
Thus, in the table we have set $o_{0123} = 1$ (so, for example, $o_{12} = o_{0123}o_{03} = o_{03}$).
}\end{Remark}

\begin{table}
\caption{The pointon properties that emerge from vanishing subspaces of spacetime tangent spaces.}
 \label{orientation table}
\begin{center}
\begin{tabular}{|l|l|l|}
\hline
free orientation & from the vanishing subspace & $\rightsquigarrow$ physical identification\\
\hline \hline
$o_0 = o_{123}$ & $e_0$ & electric charge $e^{o_0}$\\
\hdashline
$o_1 = o_{023}$ & $e_0 \wedge e_2 \wedge e_3$ & color charge $r^{o_1}$\\
$o_2 = o_{013}$ & $e_0 \wedge e_1 \wedge e_3$ & color charge $g^{o_2}$\\
$o_3 = o_{012}$ & $e_0 \wedge e_1 \wedge e_2$ & color charge $b^{o_3}$\\
\hdashline
$o_{23} = o_{01}$ & $e_2 \wedge e_3$  & spin vector $s = o_{23}e_1$\\
$o_{31} = o_{02}$ & $e_3 \wedge e_1$  & spin vector $s = o_{31}e_2$\\
$o_{12} = o_{03}$ & $e_1 \wedge e_2$  & spin vector $s = o_{12}e_3$\\
\hline
\end{tabular}
\end{center}
\end{table}

\section{Tangent space projections: spin wavefunction collapse} \label{tangent space projections}

Throughout, let $\beta$ be the worldline of a pointon.
Recall that $\beta$ is a curve in the external spacetime $\tilde{M}$ and a single point in the internal spacetime $M$.
Thus, by $\beta \subset \tilde{M}$ (resp.\ $\beta \in M$), we mean we are viewing $\beta$ as a curve in $\tilde{M}$ (resp.\ a point in $M$).
Furthermore, by $M_{\beta(t)}$ we mean the subspace of the ($4$-dimensional) tangent space $\tilde{M}_{\beta(t)}$ given by the image of $h$ at the point $\beta(t) \in \tilde{M}$; in particular, $M_{\beta(t)}$ is \textit{not} the same tangent space for each $t$.

\begin{Notation} \label{3-vector} \rm{\ 

$\bullet$ For $w \in \tilde{M}_p$, set
\begin{equation*}
\hat{h}_p(w) := \left\{ \begin{array}{ll} h_p(w)/\left|h_p(w)\right| & \text{if } \left|h_p(w)\right| \not = 0 \\ 0 & \text{otherwise} \end{array} \right.
\end{equation*}
\indent $\bullet$ We denote the parallel transport of the internal tangent space $M_{\beta(t)} \subset \tilde{M}_{\beta(t)}$ along $\beta \subset \tilde{M}$ to $\beta(t')$ by $M_{\beta(t) \to \beta(t')} \subset \tilde{M}_{\beta(t')}$. 
Specifically, $M_{\beta(t) \to \beta(t')}$ is the subspace of $\tilde{M}_{\beta(t')}$ spanned by the parallel transport of the vectors in $M_{\beta(t)}$ along $\beta \subset \tilde{M}$ (viewed as a curve in $\tilde{M}$) to the point $\beta(t')$ in $\tilde{M}$.\footnote{Note that $M_{\beta(t')}$ will in general not coincide with $M_{\beta(t) \to \beta(t')}$ whenever pointon worldlines transversely intersect at $\beta(t)$ or $\beta(t')$.
However, if pointon worldlines do not transversely intersect at $\beta(t)$ and $\beta(t')$, and $\beta$ is a timelike geodesic, then $M_{\beta(t')} = M_{\beta(t) \to \beta(t')}$.}

$\bullet$ Finally, suppose the pointon has timelike $4$-velocity $v$ and let $e_0, \ldots, e_3 \in \tilde{M}_{\beta(t)}$ be an orthonormal tetrad along $\beta \subset \tilde{M}$ for which $e_0 = v$. 
If $w, w'$ are $4$-vectors that lie in $V := \operatorname{span}\{e_1, e_2, e_3 \} \subset \tilde{M}_{\beta(t)}$, then we may write them as $3$-vectors $\bm{w}, \bm{w}' \in V$ with respect to $e_1,e_2, e_3$, and denote their contraction by the dot product in $V$, 
\begin{equation*}
\bm{w} \! \cdot \! \bm{w}' = -w^a w'_a.
\end{equation*}
}\end{Notation}

Let $\beta \subset \tilde{M}$ be a pointon worldline, and suppose that for $t \in (0, \epsilon]$, the dimensions of $M_{\beta(-t)}$ and $M_{\beta(t)}$ are constant and satisfy
\begin{equation*} \label{meet}
 \dim M_{\beta(0)} < \dim M_{\beta(\pm t)}.
\end{equation*}
Let $s$ be an \textit{internal} unit vector parallel transported along $\beta$ and set $p := \beta(0) \in \tilde{M}$.
\begin{itemize}
 \item[\textsc{(a)}] As $s$ \textit{enters} the lower dimensional internal space $M_p$, $s \in M_{\beta(-\epsilon) \to p}$ is projected onto it by the normalized internal metric $\hat{h}_p: \tilde{M}_p \to M_p$.
\item[\textsc{(b)}] As $\hat{h}_p(s)$ \textit{exits} $M_p$, the time reversal of (\textsc{a}) occurs: a unit vector section $s' \in M_{\beta(\epsilon) \to p}$ of $\hat{h}_p: \tilde{M}_p \to M_p$ is chosen, that is, 
\begin{equation} \label{equals}
\hat{h}_p(s') = \hat{h}_p(s).
\end{equation}
If $\hat{h}_p(s) \not = 0$, then the probability that $s'$ is chosen is given by what we call the \textit{Kochen-Specker conditional probability}: 
\begin{equation} \label{g(,)}
p(s' | h_p(s)) = \tfrac{1}{\pi} \hat{h}_p(\bm{s}) \! \cdot \! \bm{s}'.
\end{equation}
\end{itemize}
We thus have
\begin{equation*}
\xymatrix@R-2pc{
\tilde{M}_{\beta(-\epsilon)} & \tilde{M}_{p} \ar^{h}[rdd]
& \tilde{M}_{p} & \tilde{M}_{p}  \ar_{h}[ldd]
& \tilde{M}_{\beta(\epsilon)}\\
\cup & \cup & \cup & \cup & \cup \\
M_{\beta(-\epsilon)} \ar^{\cong \ \ \ \ \ }[r] & M_{\beta(-\epsilon) \to p} & M_{p} 
& M_{\beta(\epsilon) \to p} \ar^{\ \ \cong}[r] & M_{\beta(\epsilon)} \\
s \ar@{|->}[r] & s \ar@{|->} [r] 
\ar@{}_{\substack{ \ \\ \ \\ \text{\footnotesize{wavefunction}} \\ \text{\footnotesize{collapse}}}} [r] &
\hat{h}_p(s) = \hat{h}_p(s') \ar@{<-|} [r] 
\ar@{}_{\substack{ \ \\ \ \\ \text{ \ \ \footnotesize{randomness}} \\ 
\text{ \ \ \footnotesize{\textcolor{white}{l} appears \textcolor{white}{l}}}}} [r] & s' \ar@{|->}[r] & s' 
} 
\end{equation*}

\begin{Definition} \label{ontic} \rm{
Suppose $\dim M_p = 1$ and let $\hat{h}_p(s) = \breve{v}^a$ be the internal $4$-velocity of the pointon at $p = \beta(0)$.
We call the parallel transports of $\hat{h}_p(s)$ and $s'$ along $\beta$ to $\beta(\epsilon)$ the \textit{epistemic} and \textit{ontic spin vectors} of the pointon at $\beta(\epsilon)$, respectively.
}\end{Definition}

In this case, we propose the following:
\begin{itemize} 
 \item[(\textsc{a}$'$)] \textit{The projection $s \mapsto h_p(s)$ corresponds to a measurement of spin in the direction of the line $M_p \subset \tilde{M}_p$: a spin wavefunction collapse arises from an actual projection between tangent spaces on spacetime.}
 \item[(\textsc{b}$'$)] \textit{The choice of unit vector $s'$ corresponds to the randomness inherent in the outcome of a measurement of spin subsequent to $p = \beta(0)$.}
\end{itemize}

Suppose an ontic spin vector $s$ exits a $1$-dimensional tangent space $M_p$ at $p = \beta(0)$, is parallel transported along a pointon worldline $\beta \subset \tilde{M}$ for a time $\epsilon > 0$, and then enters another $1$-dimensional tangent space $M_q$ at $q = \beta(\epsilon)$:
\begin{equation*}
\dim M_p = \dim M_q = 1 \ \ \ \ \text{ and } \ \ \ \ \dim M_{\beta(t)} > 1 \ \text{ for } t \in (0, \epsilon).
\end{equation*}
We say the spin $s$ of the pointon is \textit{prepared} at $p$ and \textit{measured} at $q$.
Furthermore, the parallel transport of $h_p(s)$ along $\beta$ (which in general is not equal to $s$) is the pointon's epistemic spin vector, by Definition \ref{ontic}.

Let $e_0, \ldots, e_3 \in \tilde{M}_{\beta(t)}$ be an orthonormal tetrad parallel transported along $\beta$ such that $e_0 = v$. 
The $3$-vector $\hat{h}_{\beta(t)}(\bm{s}) \in \operatorname{span}\{ e_1, e_2, e_3 \} \subset \tilde{M}_{\beta(t)}$ is a unit vector, and thus may be regarded as a point on the (parallel transported) unit sphere $S^2_{\beta(t)} \subset \tilde{M}_{\beta(t)}$ defined by the triad $e_1, e_2, e_3$. 

Recall the standard correspondence between the Bloch sphere $S^2$ and the spin Hilbert space $\mathcal{H} = \mathbb{C}\left| \uparrow \right\rangle \oplus \mathbb{C}\left| \downarrow \right\rangle$ given by
\begin{equation} \label{cos}
\bm{w} = (\sin \theta \cos \varphi, \sin \theta \sin \varphi, \cos \theta) \ \longleftrightarrow \ \left| \bm{w} \right\rangle = \cos(\theta/2) \left| \uparrow \right\rangle + e^{i \varphi} \sin(\theta/2) \left| \downarrow \right\rangle,
\end{equation}
where $0 \leq \theta \leq \pi$ and $0 \leq \varphi < 2 \pi$.
Under this correspondence, we may identify $S^2_{\beta(t)}$ with the Bloch sphere, and the $3$-vector $\hat{h}_{\beta(t)}(\bm{s}) \in S^2_{\beta(t)}$ with the ket $| \hat{h}_{\beta(t)}(\bm{s}) \rangle \in \mathcal{H}$,
\begin{equation*}
h_{p}(s) \in M_p \ \ \longleftrightarrow \ \ | \hat{h}_p(\bm{s}) \rangle \in \mathcal{H} \ \ \ \ \text{ and } \ \ \ \ 
h_{q}(s) \in M_q \ \ \longleftrightarrow \ \ \langle \hat{h}_q(\bm{s}) | \in \mathcal{H}^*.
\end{equation*}
We thus obtain a spacetime geometric realization of an ontological model of spin called the Kochen-Specker model \cite{KS}, which we shall now briefly review. 

The Kochen-Specker model is a realist model where quantum spin states are epistemic: they specify our knowledge about ontic (i.e., physically real) states.
The ontic state space of the model is the Bloch sphere $S^2$.
If the spin of a particle is measured to be $\left| \bm{\psi} \right\rangle \in \mathcal{H}$, then the particle's subsequent ontic state will be some vector $\bm{\lambda} \in S^2$ that lies in the hemisphere `above' the plane orthogonal to $\bm{\psi} \in S^2$, that is, 
\begin{equation*}
\bm{\psi} \! \cdot \! \bm{\lambda} > 0.
\end{equation*}
For example, if the particle is measured to have spin up $\left| (0,0,1) \right\rangle$ (resp.\ spin down $\left| (0,0,-1) \right\rangle$) in the $z$-direction, then its subsequent ontic state is some vector that lies in the northern hemisphere (resp.\ southern hemisphere) of $S^2$. 

In this model, the probability that a preparation procedure $P_{\bm{\psi}}$ produces an ontic state $\bm{\lambda}$ represented by $\bm{\psi}$ is given by\footnote{Note that $p(\bm{\lambda} | P_{\bm{\psi}})$ is correctly normalized: it suffices to suppose  $\bm{\psi} = (0,0,1)$ and 
\begin{equation*}
\bm{\lambda} = (\sin \theta \cos \varphi, \sin \theta \sin \varphi, \cos \theta).
\end{equation*}
Then $\int_{S^2} d \bm{\lambda} \, H(\bm{\psi} \! \cdot \! \bm{\lambda}) \bm{\psi} \! \cdot \! \bm{\lambda} = \int_{-\pi}^{\pi} \int_{0}^{\pi/2} \cos \theta \sin \theta \,  d \theta d \varphi = \pi$.}
\begin{equation} \label{n}
p(\bm{\lambda} | P_{\bm{\psi}}) = \tfrac{1}{\pi} H(\bm{\psi} \! \cdot \! \bm{\lambda}) \bm{\psi} \! \cdot \! \bm{\lambda},
\end{equation}
where $H$ is the Heaviside function: $H(x) = 1$ if $x > 0$ and $0$ otherwise. 
Since the probability $p(\bm{\lambda} | P_{\bm{\psi}})$ is independent of the preparation procedure $P$, for brevity we denote it $p(\bm{\lambda} | \bm{\psi})$.
Furthermore, the probability that a measurement $\left| \bm{\phi} \right\rangle \! \left\langle \bm{\phi} \right|$ of $\bm{\lambda}$ yields $\bm{\phi} \in S^2$ is given by
\begin{equation} \label{g}
p( \bm{\phi} | \bm{\lambda}) = H(\bm{\phi} \! \cdot \! \bm{\lambda}).
\end{equation}
The conditional probabilities (\ref{n}) and (\ref{g}) together reproduce the Born rule,\footnote{For completeness, we show (\ref{p phi}) following the derivation of \cite[Appendix B]{L}.  
It suffices to suppose $\bm{\psi} = (1,0,0)$ and $\bm{\phi} = (\cos \varphi_{\bm{\phi}}, \sin \varphi_{\bm{\phi}}, 0)$.
The Heaviside functions in (\ref{p phi}) imply the restrictions $-\frac{\pi}{2} < \varphi < \frac{\pi}{2}$ and $-\frac{\pi}{2} + \varphi_{\bm{\phi}} < \varphi < \frac{\pi}{2} + \varphi_{\bm{\phi}}$.
Thus, if $\varphi_{\bm{\phi}}$ is positive, we have
\begin{multline*}
\int_{-\pi}^{\pi} \int_0^{\pi} H(\bm{\phi} \! \cdot \! \bm{\lambda} ) \frac{1}{\pi} H( \bm{\psi} \! \cdot \! \bm{\lambda}) \bm{\psi} \! \cdot \! \bm{\lambda} \sin \theta \, d\theta d \varphi = \frac{1}{\pi} \int_{-\frac{\pi}{2} + \varphi_{\bm{\phi}}}^{\frac{\pi}{2}} \int_0^{\pi} (\sin \theta \cos \varphi)  \sin\theta \, d\theta d \varphi\\ 
= \frac{1}{\pi} \frac{\pi}{2} \sin \varphi \mid_{-\frac{\pi}{2} + \varphi_{\bm{\phi}}}^{\frac{\pi}{2}} 
= \frac 12 (1 + \cos \varphi_{\bm{\phi}})
= \left| \cos \frac{\pi}{4} \cos \frac{\pi}{4} \left\langle 0 | 0 \right\rangle + e^{i \varphi_{\bm{\phi}}}\sin \frac{\pi}{4} \sin \frac{\pi}{4} \left\langle 1 | 1 \right\rangle \right|^2
\stackrel{\textsc{(i)}}{=} \left| \left\langle \bm{\phi} | \bm{\psi} \right\rangle \right|^2,
\end{multline*}
where (\textsc{i}) holds by (\ref{cos}).
A similar computation holds if $\varphi_{\bm{\phi}}$ is negative.}
\begin{equation} \label{p phi}
p(\bm{\phi} | \bm{\psi}) = \int_{S^2} d\bm{\lambda} \, p( \bm{\phi} | \bm{\lambda}) \, p( \bm{\lambda} | \bm{\psi}) = \int_{S^2} d\bm{\lambda} \, H(\bm{\phi} \! \cdot \! \bm{\lambda} ) \tfrac{1}{\pi} H( \bm{\psi} \! \cdot \! \bm{\lambda} ) \bm{\psi} \! \cdot \! \bm{\lambda} = \left| \left\langle \bm{\psi} | \bm{\phi} \right\rangle \right|^2.
\end{equation}

\begin{Theorem}
Our spacetime model of spin reproduces the Born rule for spin wavefunction collapse:
\begin{equation*}
p(h_q(s) | h_p(s)) = | \langle \hat{h}_q(\bm{s}) | \hat{h}_p(\bm{s}) \rangle |^2.
\end{equation*}
\end{Theorem}

\begin{proof}
The assumptions (\ref{equals}) and (\ref{g(,)}) of our model are equivalent to the assumptions (\ref{g}) and (\ref{n}) of the Kochen-Specker model, with
\begin{equation*}
\hat{h}_p(\bm{s}) = \bm{\psi}, \ \ \ \ \bm{s} = \bm{\lambda}, \ \ \ \ \hat{h}_q(\bm{s}) = \bm{\phi}.
\end{equation*}
Our model is therefore a spacetime geometric realization of the Kochen-Specker model, and the theorem follows.
\end{proof}

\section{Photons} \label{photons section}

In this and the following two sections we will find that \textit{a consequence of internal spacetime geometry is that any particle with a null $4$-velocity must be electrically neutral, have spin-1, and admit a trivalent interaction vertex similar to the electron-photon vertex.} 

Consider a pointon with geodesic worldline $\beta \subset \tilde{M}$ and null $4$-velocity $v$.
By Lemma \ref{null projection lemma}, a minimal orthogonal projection of $v$ is of the form $[v]_{ab} = g_{ab} + v_av'_b + v_bv'_a$ for some choice of null vector $v'$ satisfying $v^a v'_a = -1$. 
The choice of $v'$ is equivalent to a choice of unit timelike or spacelike vector
\begin{equation} \label{timelike and spacelike}
\tilde{e}_0 := \tfrac{1}{\sqrt{2}}(v - v') \ \ \ \ \text{ or } \ \ \ \ \tilde{e}_3 := \tfrac{1}{\sqrt{2}}(v + v').
\end{equation}
Hence, a choice of $v'$ is equivalent to a choice of inertial frame with time direction $\tilde{e}_0$.
The vectors $\tilde{e}_0$, $\tilde{e}_3$ are orthogonal and satisfy
\begin{equation} \label{disagree}
v = \tfrac{1}{\sqrt{2}}( \tilde{e}_0 + \tilde{e}_3) \ \ \ \ \text{ and } \ \ \ \ v' = \tfrac{1}{\sqrt{2}}(-\tilde{e}_0 + \tilde{e}_3).
\end{equation}
Thus, $\tilde{e}_3$ is the direction of propagation of the pointon in the inertial frame with time direction $\tilde{e}_0$.
Furthermore, (\ref{timelike and spacelike}) implies that there are projections
\begin{equation} \label{e_0 projection}
[v]_{ab}\tilde{e}_0^a = [v]_{ab}\tilde{e}_3^a = 0.
\end{equation}
In particular, at points $\beta(t)$ along $\beta \subset \tilde{M}$ not transversely intersected by other pointon worldlines, the internal tangent space $M_{\beta(t)} \subset \tilde{M}_{\beta(t)}$ is a plane orthogonal to $\tilde{e}_0$ and $\tilde{e}_3$. 
\textit{Consequently, there is a free choice of plane orientation $o_{03} = o(\tilde{e}_0 \wedge \tilde{e}_3)$, but no free choice of timelike orientation $o_0 = o(\tilde{e}_0)$, and therefore the pointon has no electric charge.} 

\begin{Theorem} \label{photon}
A pointon with a geodesic worldline $\beta \subset \tilde{M}$ and null $4$-velocity $v$ can only exist in a bound state with another pointon, and this state has nonzero spatial extent in the direction $v + v'$ in $\tilde{M}$. 
Furthermore, the two pointons acquire opposite electric charges in the minimal projection $[v + v']\tilde{M}_{\beta(t)} \supset M_{\beta(t)}$ of $\tilde{M}_{\beta(t)}$ where the pointons become timelike.
\end{Theorem}

\begin{proof}
Recall the timelike and spacelike unit vectors $\tilde{e}_0$ and $\tilde{e}_3$ in (\ref{timelike and spacelike}).
The vectors $v$ and $v'$ determine a unique minimal orthogonal projection of $\tilde{M}_{\beta(t)}$ for which the null pointon becomes timelike, namely 
\begin{equation*}
\tensor{[\tilde{e}_3]}{^a_b} = \tensor{[v + v']}{^a_b}.
\end{equation*}
This projection yields the inclusions
\begin{equation*}
M_{\beta(t)} \subseteq [\tilde{e}_0][\tilde{e}_3]\tilde{M}_{\beta(t)} \subset [\tilde{e}_3]\tilde{M}_{\beta(t)} \subset \tilde{M}_{\beta(t)}.
\end{equation*}
In the smaller external space $[\tilde{e}_3]\tilde{M}_{\beta(t)}$, the null pointon has a \textit{timelike} $3$-velocity $[\tilde{e}_3]v = \tilde{e}_0$, by (\ref{disagree}).
Thus, at points $\beta(t) \in \tilde{M}$ not transversely intersected by other pointon worldlines, the pointon's free choice of orientation that results from the vanishing of $\tilde{e}_0$ in $[\tilde{e}_3]\tilde{M}_{\beta(t)}$ is electric charge $o_0$, by (\ref{electric charge}).
(Note that $o_0$ and $o_{03}$ are independent since any identification would not be well defined.)
Therefore, although the pointon is neutral in $\tilde{M}_{\beta(t)}$, the pointon does have an electric charge in the smaller external space $[\tilde{e}_3]\tilde{M}_{\beta(t)}$.
But this subspace charge implies that \textit{the pointon has electric charge in the spatial plane orthogonal to $\tilde{e}_3$ in $\tilde{M}_{\beta(t)}$}, in contradiction to the pointon being neutral in $\tilde{M}_{\beta(t)}$.
This inconsistency is resolved if $\beta$ is necessarily the worldline of \textit{two} (or an even number of) pointons, both neutral in $\tilde{M}$ and with opposite electric charges in $[\tilde{e}_3]\tilde{M}_{\beta(t)}$. 

Finally, suppose that $\beta$ is not transversely intersected by other pointon worldlines at $\beta(t) \in \tilde{M}$.
Then the internal metric at $\beta(t)$ is $h = [v] = [\tilde{e}_0][\tilde{e}_3]$, by (\ref{e_0 projection}). 
Since $h$ is obtained from the external metric $g$ by projecting out tangent vectors to pointon worldlines, $\tilde{e}_3$ must be a tangent vector to the pointon bound state at $\beta(t)$.
Consequently, the bound state has nonzero spatial extent in the $\tilde{e}_3$ direction in $\tilde{M}$.
More precisely, in the inertial frame with time direction $\tilde{e}_0$, there is some $\epsilon >0$ for which the bound state contains the locus $\exp_{\beta(t)}( (-\epsilon, \epsilon) \tilde{e}_3) \subset \tilde{M}$. 
\end{proof}

\begin{figure}
\includegraphics[width=14.3cm]{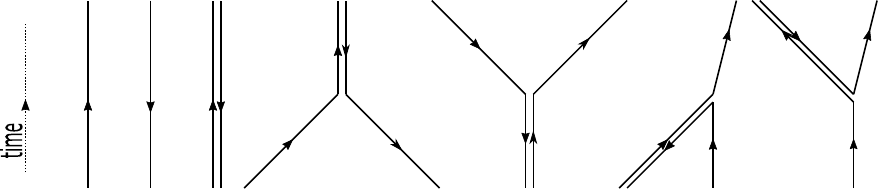}
\caption{From left to right: an electron; a positron; a photon; an electron and positron join to form a photon; a photon splits into an electron and positron; an electron absorbs a photon; and an electron emits a photon. The arrowheads indicate time orientations.}
\label{fig1}
\end{figure}

We therefore propose that \textit{bound pairs of pointons of opposite electric charge are photons, and free pointons are electrons or positrons, depending on their charge}.
Furthermore, by allowing pairs of pointons of opposite charge to be created or annihilated, we obtain the electron-photon trivalent vertex.
This is shown in Figure \ref{fig1}.\footnote{Our composite model of electrons and photons is similar, in a sense, to 't Hooft's double line formalism for quark-gluon interactions \cite{'tH}.} 
We will henceforth refer to bound pairs of pointons of opposite charge as `photons' and free pointons as `electrons'.

The constituent pointons of a photon each possess a spin $4$-vector; we call these vectors $s_1$ and $s_2$.
In the next section we will use these vectors to model spin-$1$ polarization.

We defined a spin vector $s$ exiting a $1$-dimensional tangent space to have \textit{unit length}, $|s^2| = 1$, since spin vectors result from a free choice of plane orientation and an orientation does not specify a length. 
Similarly, we define photon spin vectors $s_1, s_2$ exiting a $1$- or $2$-dimensional tangent space to span a parallelogram of \textit{unit area}, since two plane orientations do not specify an area.  
But if two vectors have unit length and span a parallelogram with unit area, then they are orthogonal.
Consequently, photon spin vectors $s_1, s_2$ are orthogonal,
\begin{equation} \label{orthogonal s}
\tensor{s}{_1^a} \tensor{s}{_{2a}} = 0.
\end{equation}

For a photon with null $4$-velocity $v$, a choice of null vector $v'$---which is required by Lemma \ref{null projection lemma}---will serve two purposes in our model:
\begin{itemize}
 \item[(i)] In Section \ref{polarization section} we will show that $v'$ uniquely determines the photon's polarization type: linear, elliptical, or circular.
 \item[(ii)] Although the photon has a null $4$-velocity, $v'$ specifies an inertial frame from the relations (\ref{timelike and spacelike}). 
This inertial frame will be used to define the energy $\omega > 0$ and $4$-momentum $k = \omega v$ of the photon in \cite{B7}.
\end{itemize}

\section{Pairs of spin vectors: photon polarization} \label{polarization section}

In Section \ref{photons section} we introduced a composite photon-electron model where a photon consists of two pointons, one with positive timelike orientation (an electron) and one with negative timelike orientation (a positron).
Attached to each pointon is a spin vector, and thus a photon has two spin vectors.
In this section we will introduce an extension of the Kochen-Specker model to spin-$1$ polarization and its wavefunction collapse, by using pairs of spin vectors in place of single spin vectors.
We assume that photons travel with speed $|\bm{v}|$ equal to $c$ in a vacuum and less than $c$ in a medium, as is the case for electromagnetic waves (e.g., \cite[Chapt.\ 1]{BW}).\footnote{In a homogeneous isotropic medium with dielectric constant $\epsilon$ and magnetic permeability $\mu$, we assume $|\bm{v}| = c/\sqrt{\epsilon \mu}$, in accordance with Snell's law.
Of course, the phase velocity or group velocity of a \textit{superposition} of monochromatic electromagnetic waves, all with speeds less than $c$, may exceed $c$ or be negative (as in anomalous dispersion).}

\subsection{Photon polarization in a medium} \label{medium}

Consider a photon with worldline $\beta \subset \tilde{M}$, timelike $4$-velocity $v$, and spin $4$-vectors $s_1, s_2$.
We would like to identify the pair $s_1,s_2$ with the polarization of the photon.

Let $e_0, \ldots, e_3 \in \tilde{M}_{\beta(t)}$ be an orthonormal tetrad parallel transported along $\beta \subset \tilde{M}$ such that $v = \tfrac{1}{\sqrt{1 - a^2}}(e_0 + a e_3)$ for some $a \in (0,1)$.
Since $v^2 \not = 0$, $h_{\beta(t)}$ projects out the $1$-dimensional subspace of $\tilde{M}_{\beta(t)}$ spanned by $v$ at points $\beta(t)$ along $\beta$.
We will consider the $3$-vectors $\bm{s}_j := [e_0]s_j \in \operatorname{span}\{e_1,e_2,e_3\} \subset \tilde{M}_{\beta(t)}$ in the reference frame with (positive) time direction $e_0$.

We briefly recall the classical and quantum descriptions of polarization.
In the reference frame with time direction $e_0$, a monochromatic electromagnetic plane wave with Poynting vector $\mathbf{S} = |\mathbf{S}|e_3$ has the form
\begin{equation*} \label{mono}
\mathbf{E} = \operatorname{Re}(E_1 e^{ik_ax^a}) e_1 + \operatorname{Re}(E_2 e^{ik_ax^a})e_2,
\end{equation*}
where $E_1,E_2 \in \mathbb{C}$ are complex amplitudes and $\operatorname{Re}$ denotes the real part. 
The ratio of the amplitudes characterizes the type of polarization of $\mathbf{E}$:
\begin{equation*}
\frac{E_2}{E_1} = \left\{ 
\begin{array}{lcl} 
\pm \frac{|E_2|}{|E_1|} & \Leftrightarrow & \text{ linear with $E_1 \not = 0$} \\
 -i & \Leftrightarrow & \text{ right-handed circular} \\
 i & \Leftrightarrow & \text{ left-handed circular} 
\end{array} \right.
\end{equation*}
Furthermore, $E_2/E_1$ has negative (resp.\ positive) imaginary part if and only if $\mathbf{E}$ has right-handed (resp.\ left-handed) elliptical polarization. 
In general, the polarization of $\mathbf{E}$ may be represented by its Jones vector
\begin{equation*}
\psi = \left[ \begin{matrix} E_1 \\ E_2 \end{matrix} \right] \in \mathcal{H} \cong \mathbb{C}^2
\end{equation*}
in the (quantum) Hilbert space $\mathcal{H}$. 
Set 
\begin{equation*}
\left| e_1 \right\rangle := \left[ \begin{smallmatrix} 1 \\ 0 \end{smallmatrix} \right], \ \ \ \ \
\left| e_2 \right\rangle := \left[ \begin{smallmatrix} 0 \\ 1 \end{smallmatrix} \right].
\end{equation*}

To associate a polarization state to the pair $s_1, s_2$, decompose each $\bm{s}_j = [e_0]s_j$ into $3$-vectors parallel and orthogonal to the photon's direction of propagation $e_3$,
\begin{equation*}
\bm{s}_j = \bm{s}_j^{\parallel} + \bm{s}_j^{\perp},
\end{equation*}
where $\bm{s}_j^{\|} := (\bm{s}_j \! \cdot \! e_3)e_3$.
Note that by our choice of tetrad, namely $v = e_0 + a e_3$, $s_j$ has $e_0$ component $a (s_j \! \cdot \! e_3)$, since $s_j^a v_a = 0$ by assumption.

Since $\bm{s}_1 \! \cdot \! \bm{s}_2 = 0$, at least one of $\bm{s}_1^{\perp}, \bm{s}_2^{\perp}$ is nonzero; say $\bm{s}_1^{\perp} \not = 0$.
Let $\theta \in [0, 2 \pi)$ be such that 
\begin{equation*}
\hat{\bm{s}}_1^{\perp} := \frac{\bm{s}_1^{\perp}}{|\bm{s}_1^{\perp}|} = \cos \theta  e_1 + \sin \theta e_2.
\end{equation*}
If $\bm{s}_2^{\perp} = 0$, that is, $\bm{s}_2 = \bm{s}_2^{\|}$ is parallel to $e_3$, then $\bm{s}_1 \! \cdot \! \bm{s}_2 = 0$ implies that $\bm{s}_1 = \bm{s}_1^{\perp}$.
We thus define
\begin{equation} \label{s_2 perp}
\hat{\bm{s}}_2^{\perp} := \left\{ \begin{array}{ll} \frac{\bm{s}_2^{\perp}}{| \bm{s}_2^{\perp}|} & \text{ if } \bm{s}_2^{\perp} \not = 0\\
\sin \theta e_1 - \cos \theta e_2 & \text{ otherwise } \end{array} \right.
\end{equation}
We map the spin vectors $s_1,s_2$ to a monochromatic electromagnetic plane wave $\mathbf{E}$ with complex amplitudes
\begin{align*}
E_j & = \tfrac{1}{\sqrt{2}} \! \left[ (| \bm{s}_1^{\perp}| + i | \bm{s}_1^{\|}| ) \hat{\bm{s}}_1^{\perp} + (| \bm{s}_2^{\perp}| + i | \bm{s}_2^{\|}| )\hat{\bm{s}}_2^{\perp} \right] \! \! \cdot \! e_j\\
& = \tfrac{1}{\sqrt{2}} \! \left[ (\bm{s}_1 + \bm{s}_2)^{\perp} + i(e_3 \! \cdot \! \bm{s}_1 \hat{\bm{s}}_1^{\perp} + e_3 \! \cdot \! \bm{s}_2 \hat{\bm{s}}_2^{\perp}) \right] \! \! \cdot \! e_j, 
\end{align*}
for $j = 1,2$.
Equivalently, setting
\begin{equation*}
| \hat{\bm{s}}_j^{\perp} \rangle := \hat{\bm{s}}_j^{\perp} \! \cdot \! e_1 | e_1 \rangle + \hat{\bm{s}}_j^{\perp} \! \cdot \! e_2 | e_2 \rangle \in \mathcal{H},
\end{equation*}
we map $s_1,s_2$ to the Jones vector
\begin{equation*}
\psi = \tfrac{1}{\sqrt{2}} \! \left[ (|\bm{s}_1^{\perp}| + i |\bm{s}_1^{\|}| ) | \hat{\bm{s}}_1^{\perp} \rangle + (|\bm{s}_2^{\perp}| + i |\bm{s}_2^{\|}| ) | \hat{\bm{s}}_2^{\perp} \rangle \right].
\end{equation*}

Under these identifications we recover exactly the classical and quantum descriptions of linear and circular polarization:

\begin{Lemma} \label{polarization map}
Consider a timelike photon with spin vectors $s_1, s_2$.
\begin{enumerate}
 \item Suppose $\bm{s}_1 = \bm{s}_1^{\parallel} = \pm e_3$ or $\bm{s}_2 = \bm{s}_2^{\parallel} = \pm e_3$.
Then $E_2/E_1 = \pm i$, and thus the photon has circular polarization.
 \item Suppose $\bm{s}_1 = \bm{s}_1^{\perp}$ and $\bm{s}_2 = \bm{s}_2^{\perp}$.
Then $E_2/E_1 = \pm |E_2|/|E_1|$, and thus the photon has linear polarization.
\end{enumerate}
\end{Lemma}

\begin{proof} 
(1) First suppose $\bm{s}_1 = \pm e_3$.
Then $\bm{s}_2 = \bm{s}_2^{\perp} = \hat{\bm{s}}_2^{\perp}$ since $\bm{s}_1 \! \cdot \! \bm{s}_2 = 0$.
Thus, the amplitudes $E_j$ reduce to 
\begin{equation*}
E_j = \tfrac{1}{\sqrt{2}}(\hat{\bm{s}}_1^{\perp} + i \hat{\bm{s}}_2^{\perp}) \! \cdot \! e_j.
\end{equation*}
Consequently, (\ref{s_2 perp}) implies that for some $\theta \in [0, 2 \pi)$,
\begin{align*}
E_1 & = \tfrac{1}{\sqrt{2}}( \cos \theta \pm i \sin\theta)\\
E_2 & = \tfrac{1}{\sqrt{2}}(\sin \theta \mp i \cos \theta)\\
& = \mp i \tfrac{1}{\sqrt{2}}( \cos \theta \pm i \sin \theta)\\
& = \mp i E_1.
\end{align*}
Whence, $E_2/E_1 = \mp i$.

(2) Now suppose $\bm{s}_1 = \bm{s}_1^{\perp}$ and $\bm{s}_2 = \bm{s}_2^{\perp}$, that is, $\bm{s}_1$ and $\bm{s}_2$ both lie in the plane orthogonal to the photon's direction of propagation $e_3$.
Then the amplitudes $E_j$ reduce to
\begin{equation*}
E_j = \tfrac{1}{\sqrt{2}} (\bm{s}_1+\bm{s}_2) \! \cdot \! e_j. 
\end{equation*}
Therefore, since $e_1 \! \cdot \! e_2 = 0$,
\begin{equation*}
\frac{E_2}{E_1} = \frac{(\bm{s}_1 + \bm{s}_2) \! \cdot \! e_2}{(\bm{s}_1 + \bm{s}_2) \! \cdot \! e_1} = \frac{\cos (\theta + \pi/2)}{\cos \theta} = - \tan \theta = \pm \frac{|E_2|}{|E_1|}.
\end{equation*}
\end{proof}

Similarly, letting $|H \rangle := | e_1 \rangle$ and $| V \rangle := | e_2 \rangle$ we obtain exactly the Jones vectors for the horizontal-vertical, diagonal-antidiagonal, and circular polarization bases of $\mathcal{H}$:
\begin{align} \label{nguoakz}
\begin{split}
\left| \bm{s}_1 = e_1 - e_2, \, \bm{s}_2 = e_1 + e_2 \right\rangle & = \left| H \right\rangle\\
\left| \bm{s}_1 = e_1+e_2, \, \bm{s}_2 = -e_1 + e_2 \right\rangle & = \left| V \right\rangle\\
\left| \bm{s}_1 = e_1, \, \bm{s}_2 = e_2 \right\rangle & = \tfrac{1}{\sqrt{2}}\left(\left| H \right\rangle + \left| V \right\rangle \right)  = \left| D \right\rangle\\
\left| \bm{s}_1 = -e_2, \, \bm{s}_2 = e_1 \right\rangle &  = \tfrac{1}{\sqrt{2}}\left(\left| H \right\rangle - \left| V \right\rangle \right) = \left| A \right\rangle\\
\left| \bm{s}_1 = e_1, \, \bm{s}_2 = e_3 \right\rangle & = \tfrac{1}{\sqrt{2}}\left(\left| H \right\rangle + i \! \left| V \right\rangle \right) = \left| L \right\rangle\\
\left| \bm{s}_1 = -e_3, \, \bm{s}_2 = e_1 \right\rangle & = \tfrac{1}{\sqrt{2}}\left(\left| H \right\rangle - i \! \left| V \right\rangle \right) = \left| R \right\rangle
\end{split}
\end{align}
Note that the factor of $i$ in the circular polarizations $| L \rangle, | R \rangle$ corresponds to a spin vector parallel to the photon's direction of propagation, as in Lemma \ref{polarization map}.

Recall that a waveplate is an optical device that may be used to change linear polarization to circular polarization.
For a photon with spin vectors $\bm{s}_1, \bm{s}_2$ incident on a waveplate, we propose that the components ${\bm{s}_j \! \cdot \! e_{\text{slow}}}$ of the spin vectors $\bm{s}_j$ along the slow axis $e_{\text{slow}}$ remain unchanged, whereas the components ${\bm{s}_j \! \cdot \! e_{\text{fast}}}$ of the spin vectors $\bm{s}_j$ along the fast axis $e_{\text{fast}}$ are rotated in the plane orthogonal to the slow axis $e_{\text{slow}}$.
In order for our pointon model to agree with the electromagnetic wave model, we take the rotation angles to be the relative phase shifts of the corresponding electromagnetic wave $\mathbf{E}$. 

Specifically, consider a photon propagating in the $e_3$ direction with initial spin vectors $\bm{s}_1 = e_1$ and $\bm{s}_2 = e_2$, that is, with diagonal polarization $| D \rangle$, incident on a waveplate with slow axis $e_1$ and fast axis $e_2$.
\begin{itemize}
 \item[(i)] If the waveplate is a quarter plate, then $\bm{s}_2$ will be rotated $90^{\circ}$ in the $e_2 \wedge e_3$ plane, whence $\bm{s}_2$ will exit pointing in the $e_3$ direction, while $\bm{s}_1$ will exit unchanged.
The photon will thus exit with circular polarization.
 \item[(ii)] If the waveplate is a half plate, then $\bm{s}_2$ will be rotated $180^{\circ}$ in the $e_2 \wedge e_3$ plane, so $\bm{s}_2$ will exit pointing in the $-e_2$ direction, while again $\bm{s}_1$ will exit unchanged.
The photon will therefore exit with antidiagonal polarization $| A \rangle$.
\end{itemize}

\begin{Theorem} \label{orthogonal lemma}
Let $\{s_1, s_2\}$ and $\{s'_1,s'_2\}$ be two polarizations. 
Then, by possibly relabeling subscripts,
\begin{equation*}
\left\langle s_1, s_2 \mid s'_1, s'_2 \right\rangle = 0 \ \ \ \ \Longleftrightarrow \ \ \ \ s_1 = s'_1 \text{ and } s_2 = -s_2'. \end{equation*}
That is, the quantum state vectors of two polarizations are orthogonal in the Hilbert space $\mathcal{H} \cong \mathbb{C}^2$ if and only if the pairs of spin $4$-vectors are parallel and antiparallel in spacetime.
\end{Theorem}

\begin{proof}
This is readily verified from the basis vectors (\ref{nguoakz}).
\end{proof}

\begin{Remark} \rm{ 
There are differences between our composite model of photon polarization and the electromagnetic wave and Jones vector models.

(i) In the electromagnetic wave model, the electric vector $\mathbf{E}$ is always transverse to the direction of wave propagation, $\mathbf{S} = \mathbf{E} \times \mathbf{H}$.
In contrast, the spin vectors $\bm{s}_1, \bm{s}_2$ need not be orthogonal to the direction of propagation of the photon: by Lemma \ref{polarization map}, circular polarization is obtained by $\bm{s}_1$ or $\bm{s}_2$ being parallel to the photon's direction of propagation.
Nevertheless, the spin \textit{$4$-vectors} $s_1, s_2$ are always orthogonal to the photon's \textit{$4$-velocity} $v$: $s^a_1 v_a = s^a_2 v_a = 0$. 
This is compatible with the Lorenz gauge $\partial_a A^a = 0$ together with Maxwell's equation $\partial_a F^{ab} = 0$, where $F_{ab} := \partial_a A_b - \partial_b A_a$. 

(ii) In Section \ref{tangent space projections} we showed that the Bloch sphere for electron spin is a geometric feature of internal spacetime geometry. 
In contrast, the Poincar\'e sphere for photon polarization does \textit{not} correspond to physical spacetime geometry. 
Indeed, orthogonal polarization states are represented by antipodal points on the Poincar\'e sphere (similar to the Bloch sphere), but Lemma \ref{orthogonal lemma} implies that this is not the case for the (parallel transported) unit sphere that parameterizes the possible vectors $\tfrac{1}{\sqrt{2}}( \bm{s}_1 + \bm{s}_2 )$.
}\end{Remark}

\subsection{Photon polarization in a vacuum} \label{null photon section}

Consider a photon with geodesic worldline $\beta \subset \tilde{M}$, null $4$-velocity $v$, and spin $4$-vectors $s_1, s_2$.
Since $v^2 = 0$, Lemma \ref{null projection lemma} implies that $h_{\beta(t)}$ projects out a $2$-dimensional subspace of $\tilde{M}_{\beta(t)}$ at points $\beta(t)$ along $\beta$ where the photon is isolated:
\begin{equation*}
h_{ab} = g_{ab} + v_av'_b + v_bv'_a,
\end{equation*}
with $v'$ some null $4$-vector satisfying $v^av'_a = -1$.
In the following, we describe the constraints that the null condition $v^2 = 0$ imposes on $s_1, s_2$, and thus on the photon's polarization. 

\begin{Proposition} \label{null photon prop}
Let $e_0, \ldots, e_3 \in \tilde{M}_{\beta(t)}$ be an orthonormal tetrad parallel transported along $\beta \subset \tilde{M}$ for which
\begin{equation*}
v = e_0 + e_3 = (1,0,0,1) \ \ \ \ \text{ and } \ \ \ \ s_1 = e_1  = (0,1,0,0).
\end{equation*}
Then for any choice of $\alpha, \beta, \gamma \in \mathbb{R}$ satisfying $\alpha^2 + \beta^2 + \gamma^2 = 1$ and $\gamma \not = -1$, the $4$-vector
\begin{equation*}
v' := \tfrac{1}{1 - \gamma}(-1, \alpha, \beta, \gamma)
\end{equation*}
satisfies $v'^2 = 0$ and $v^a v'_a = -1$.
The pair $v,v'$ may therefore be associated to a null photon.
Furthermore, the following holds:
\begin{enumerate}
 \item If $v' = (a, 0, b, c)$, then
\begin{equation*}
c = - a - b^2 \ \ \ \ \text{ and } \ \ \ \ s_2 = (-c, 0, 1, -c).
\end{equation*} 
 \item If $s_2 = (a, 0, b, c)$, then
\begin{equation*}
c = a  \ \ \ \ \text{ and } \ \ \ \ v' = \left(-\tfrac 12 (1+ \tfrac{a^2}{b^2}), 0, -\tfrac{a}{b}, \tfrac 12 (1 - \tfrac{a^2}{b^2} ) \right).
\end{equation*}
In particular, $b \not = 0$.
\item If $s_1, s_2$ are possible spin vectors for some $v,v'$, then so are 
\begin{equation*}
s'_1 := \cos \theta s_1 + \sin \theta s_2 \ \ \ \ \text{ and } \ \ \ \ s'_2 := \sin \theta s_1 - \cos \theta s_2
\end{equation*}
for any $\theta \in [0, 2\pi )$.
\end{enumerate}
\end{Proposition}

\begin{proof}
Follows from the relations $v^av'_a = -1$ and $v^2 = v'^2 = s_j^a v_a = s_j^av'_a = 0$. 
\end{proof}

\begin{Example} \rm{
Suppose $v = (1, 0, 0, 1)$ and $s_1 = (0,1, 0, 0)$.
Then three possible choices of $v'$ and $s_2$ are
\begin{align} 
v' = \left( -\tfrac 12, 0, 0, \tfrac 12 \right), \ \ \ \ & s_2 = (0, 0, 1, 0); \label{first ex}\\ 
v' = ( -1, 0, 1, 0 ), \ \ \ \ & s_2 = (-1, 0, 1, -1); \label{second ex}\\
v' = ( \sqrt{2} -2, 0 , \sqrt{2} - 1, \sqrt{2} -1 ), \ \ \ \ & s_2 = ( 1- \sqrt{2}, 0, 1, 1 - \sqrt{2} ). \nonumber
\end{align}
In the first example (\ref{first ex}), the polarization is linear in the $e_1 \wedge e_2$ plane, by Lemma \ref{polarization map}.
Furthermore, the photon's `intrinsic' time direction $\tilde{e}_0$ and propagation direction $\tilde{e}_3$ are related to $e_0$ and $e_3$ in the first example (\ref{first ex}) by
\begin{equation*}
\tilde{e}_0 = \tfrac{1}{\sqrt{2}}(\tfrac 32 e_0 + \tfrac 12 e_3) = \tfrac{1}{\sqrt{2}}(\tfrac 32 , 0, 0, \tfrac 12) \ \ \ \ \text{ and } \ \ \ \ \tilde{e}_3 = \tfrac{1}{\sqrt{2}}(\tfrac 12 e_0 + \tfrac 32 e_3) = \tfrac{1}{\sqrt{2}}(\tfrac 12, 0, 0, \tfrac 32).
\end{equation*}
Similarly, in the second example (\ref{second ex}) we have
\begin{equation*}
\tilde{e}_0 = \tfrac{1}{\sqrt{2}}(2,0,-1,1) \ \ \ \ \text{ and } \ \ \ \ \tilde{e}_3 = \tfrac{1}{\sqrt{2}}(0,0,1,1).
\end{equation*}
It is easy to check in both cases that $\tilde{e}_0$ and $\tilde{e}_3$ are unit vectors and $s_j^a \tilde{e}_{0a} = 0 = s_j^a \tilde{e}_{3a}$.
}\end{Example}

\begin{Corollary}
A null photon cannot possess exact circular polarization, but only an approximate circular polarization where one of its two spin $3$-vectors $\bm{s}_j := [e_0] s_j$ points arbitrarily close to the direction of the photon's propagation, up to sign.
\end{Corollary}

\begin{proof}
Follows from Proposition \ref{null photon prop}.2, since $b$ is nonzero but can be chosen to be arbitrarily close to zero.
\end{proof}

\section{Polarization wavefunction collapse}

Consider a photon with worldline $\beta \subset \tilde{M}$, timelike or null $4$-velocity $v$, and spin $4$-vectors $s_1, s_2$.
Suppose the  photon meets an electron at a point $p \in \tilde{M}$. 
In our composite model of the standard model \cite{B1,B2}, we declare there to be an electron-photon trivalent vertex at $p$ if and only if
\begin{equation} \label{polarization condition}
h_p(s_1) \left| h_p(s_2) \right| = -h_p(s_2) \left| h_p(s_1) \right|.
\end{equation}
This simplifies to $\hat{h}_p(s_1) = -\hat{h}_p(s_2)$ whenever $h_p(s_1)$ and $h_p(s_2)$ are nonzero.
If we omit the minus sign in (\ref{polarization condition}), then the bound state of the two pointons represents the $Z$ boson instead of the photon; see \cite[Table 1]{B2}. 
Note the similarity between (\ref{polarization condition}) and (\ref{equals}).

\begin{Lemma} \label{on shell}
Suppose there is an electron-photon vertex at $p \in \tilde{M}$ which conserves $4$-momentum.
Let $v$ be the photon's $4$-velocity.
Then generically,
\begin{equation} \label{dimension =}
\dim M_p = \left\{ \begin{array}{ll}
1 & \text{if $v$ is null}\\
2 & \text{if $v$ is timelike}
\end{array} \right.
\end{equation}
\end{Lemma}

\begin{proof}
Let $v_1, v_2,v_3 \in \tilde{M}_p$ be the respective $4$-velocities of the photon and two electrons at $p$, and denote by $k_i = \omega_i v_i$,  $\omega_i > 0$, their $4$-momenta.
By $4$-momentum conservation we have
\begin{equation*} \label{incoming}
\sum_{\text{incoming}} \omega_i v_i = \sum_{\text{incoming}} k_i = \sum_{\text{outgoing}} k_j = \sum_{\text{outgoing}} \omega_j v_j.
\end{equation*}
The set $v_1, v_2, v_3$ is thus linearly dependent. 
The internal metric $h$ is therefore given by the composition of projections
\begin{equation*}
h = \tensor{h}{^a_b} = \tensor{[v_1]}{^a_c}\tensor{[v_2]}{^c_b}.
\end{equation*}
Furthermore, the rank of the projection (\ref{projection of v}) of a timelike vector is $3$, and the rank of the projection (\ref{null projection}) of a null vector is $2$.
Consequently, if $v_1, v_2$ are generic, $v_1 \not = -v_2$, then (\ref{dimension =}) holds.
\end{proof}

We now show that our model yields a partial ontological spacetime description of linear polarizers.

Consider a photon with normal incidence on a linear polarizer. 
Suppose the polarizer is at rest (with $4$-velocity $e_0$) and the photon has initial $4$-velocity 
\begin{equation*}
v = \tfrac{1}{\sqrt{1 - a^2}}(e_0 + a e_3)
\end{equation*}
for some $a \in (0,1]$.
Further suppose that, at a point $p \in \tilde{M}$ in the polarizer, the photon meets an electron that is confined to a straight wire (or molecule) which runs along the $\pm e_1$ direction.
The wire thus acts as an external constraint to the photon-electron interaction at $p$.
In particular, Lemma \ref{on shell} does not hold in this case.
Consequently, for some $b, b' \in (0,1)$, the electron has ideal initial $4$-velocity $\tfrac{1}{\sqrt{1 - b^2}}(e_0 + b e_1)$ and final $4$-velocity $\tfrac{1}{\sqrt{1 - b'^2}}(e_0 + b' e_1)$. 
Whence,
\begin{equation*}
h_p(e_0 + a e_3) = h_p( e_0 + b e_1) = h_p (e_0 + b' e_1) = 0.
\end{equation*}
Thus, if $b \not = b'$, then
\begin{equation*}
h_p(e_0) = h_p(e_1) = h_p(e_3) = 0,
\end{equation*}
and so the only surviving direction is $e_2$: $M_p = \operatorname{span}\{e_2\} \subset \tilde{M}_p$.

If there is an electron-photon vertex at $p$, then we say the photon is `absorbed' by the electron; otherwise the photon is `transmitted' at $p$. 
\begin{itemize}
 \item[(i)] If the photon has circular polarization, then one of its spin vectors $s_j$ is $\pm e_3$.
Whence, 
\begin{equation*}
h_p(s_j) = h_p(e_3) = 0.
\end{equation*}
Thus, (\ref{polarization condition}) trivially holds.
Therefore \textit{a photon with circular polarization will be absorbed by the electron.}
 \item[(ii)] If the photon has linear polarization, then (\ref{orthogonal s}) and (\ref{polarization condition}) together imply that the photon is transmitted at $p$ if and only if 
\begin{equation} \label{ss<}
|\angle (s_1 + s_2, e_2)| < \tfrac{\pi}{4}.
\end{equation}
Indeed, if $|\angle (s_1 + s_2, e_2)| = \tfrac{\pi}{4}$, then the photon has diagonal or antidiagonal polarization,
\begin{equation*}
s_1 + s_2 = \pm (e_1 + e_2) \ \ \ \ \text{ or } \ \ \ \ s_1 + s_2 = \pm (e_1 - e_2).
\end{equation*}
(\ref{orthogonal s}) then implies that one of the spin vectors $s_j$ is $\pm e_1$.
Thus,
\begin{equation*}
h_p(s_j) = h_p(e_1) = 0.
\end{equation*}
But then (\ref{polarization condition}) trivially holds.
Therefore the photon is absorbed by the electron.
\item[(iii)] If the electron emits a new photon, then the photon's spin vectors will satisfy (\ref{ss<}) by (\textsc{b}) in Section \ref{tangent space projections}.
\end{itemize}

A photon may propagate \textit{between} the wires (or stretched molecules) in a linear polarizer, yet still exit with linear polarization.
We expect that this may be dealt with using pointon spinors, introduced in \cite[Section 3]{B2}, though this requires further investigation. 

\appendix
\section{Internal tangent spaces} \label{tangent space definitions}

\indent In the following, we contrast tangent spaces on a manifold with tangent spaces on an internal spacetime. 

Following the notation of \cite{dC}, recall that a (differentiable) manifold of dimension $n$ is a set $M = M^n$ and a family of injective maps $\mathbf{x}_{\alpha}: U_{\alpha} \subseteq \mathbb{R}^n \to M$ on open sets $U_{\alpha}$ such that 
\begin{itemize}
 \item[(i)] $\cup_{\alpha} \mathbf{x}_{\alpha}(U_{\alpha}) = M$; 
 \item[(ii)] for any $\alpha, \beta$ with $\mathbf{x}_{\alpha}(U_{\alpha}) \cap \mathbf{x}_{\beta}(U_{\beta})  = V \not = \emptyset$, the sets $\mathbf{x}_{\alpha}(V)$ and $\mathbf{x}_{\beta}(V)$ are open in $\mathbb{R}^n$ with $\mathbf{x}_{\beta}^{-1} \mathbf{x}_{\alpha}$ differentiable; and 
 \item[(iii)] the family $\{ (U_{\alpha},\mathbf{x}_{\alpha}) \}$ is maximal with respect to (i) and (ii).
\end{itemize}
For a point $p \in M$, a pair $(U_{\alpha}, \mathbf{x}_{\alpha})$ with $p \in \mathbf{x}_{\alpha}(U_{\alpha})$ is called a parametrization of $M$ at $p$.

Now consider an internal spacetime $M$ with a single pointon worldline $\beta: (-\epsilon, \epsilon) \to \tilde{M}$. 
Set $p := \beta(0) \in \tilde{M}$.
Let $(U, \mathbf{x})$ be a parametrization of the external spacetime $\tilde{M}$ at $p$. 
Consider a map 
\begin{equation} \label{map y}
\mathbf{y}: U \to M
\end{equation}
with $\mathbf{y} \mathbf{x}^{-1}(\beta(0)) = \beta \in M$ and continuous on the restriction $U\setminus \mathbf{x}^{-1}(\beta(-\epsilon, \epsilon))$.
Observe that $\mathbf{y}$ cannot be injective since 
\begin{equation*}
\mathbf{y} \mathbf{x}^{-1}(\beta(t)) = \mathbf{y} \mathbf{x}^{-1}(\beta(0))
\end{equation*}
for all $t \in (-\epsilon, \epsilon)$.
Furthermore, $\mathbf{y}$ cannot be continuous since any open set $V \subset M$ containing $\beta \in M$ (with the induced topology from $\tilde{M}$) will have a preimage that is not open. 
Therefore, $M$ is not a manifold.

Let us also briefly recall the definition of tangent space:
\begin{itemize}
 \item A mapping $f: M^m_1 \to M^n_2$ between manifolds $M_1$, $M_2$ is said to be differentiable at $p \in M_1$ if, for each parametrization $\mathbf{y}: V \subseteq \mathbb{R}^n \to M_2$ at $f(p)$, there is a parametrization $\mathbf{x}: U \subseteq \mathbb{R}^m \to M_1$ at $p$ such that $f(\mathbf{x}(U)) \subseteq \mathbf{y}(V)$ and 
\begin{equation*}
\mathbf{y}^{-1} \, f \, \mathbf{x}: U \subseteq \mathbb{R}^m \to \mathbb{R}^n
\end{equation*}
is differentiable at $\mathbf{x}^{-1}(p)$.
 \item Let $\gamma: (-\epsilon, \epsilon) \to M$ be a differentiable curve in a manifold $M$ and set $p := \gamma(0)$.
Let $\mathcal{D}$ be the set of functions $f: M \to \mathbb{R}$ that are differentiable at $p$.
Then the tangent vector to $\gamma$ at $t = 0$ is the function $\gamma'(0): \mathcal{D} \to \mathbb{R}$ defined on $f \in \mathcal{D}$ by
\begin{equation} \label{usual definition}
\gamma'(0)f := \left. \frac{d((f \mathbf{x})(\mathbf{x}^{-1}\gamma))}{dt} \right\rvert_{t= 0} = \left. \frac{d(f \gamma)}{dt} \right\rvert_{t= 0}.
\end{equation}
 \item Finally, the set of all such tangent vectors is called the tangent space $M_p$ of $M$ at $p$.
\end{itemize}

The problem with applying these definitions to an internal spacetime $M$ is that $\mathbf{y}: U \to M$ in (\ref{map y}) is not injective.
Thus, we cannot replace $\mathbf{x}^{-1}$ in (\ref{usual definition}) with $\mathbf{y}^{-1}$.
Morally, the tangent vector $v = \beta'(0)$ to a pointon worldline $\beta$ should be zero on $M$ because $\beta: (-\epsilon, \epsilon) \to M$ is a constant map, and so any directional derivative along $\beta$ should vanish.
However, without a parametrization of $M$ at $\beta \in M$ we cannot make this precise using (\ref{usual definition}).
To formulate a suitable definition of tangent space $M_p$ of $M$ at $p \in \tilde{M}$, we require the following:\footnote{To note, $M_p$ is defined at a point $p$ of $\tilde{M}$, \textit{not of $M$}.}
\begin{itemize}
 \item[(i)] $M_p$ should be a vector space,\footnote{If we simply removed the vector $v$ from $\tilde{M}_p$, then the resulting set $\tilde{M}_p \setminus \{ v\}$ would not be a vector space and its elements would not be vectors. 
Whence, we would loose all notion of `tangent vector'.} and in particular a subspace of $\tilde{M}_p$; and 
 \item[(ii)] $M_p$ should not contain tangent vectors to pointon worldlines at $p$.
\end{itemize}

These two conditions naturally lead to orthogonal projections.
Indeed, suppose that there is a single timelike pointon worldline $\beta$ in $\tilde{M}$ at $p = \beta(0)$.
Fix an orthonormal basis of $\tilde{M}_p$ that contains $v = \beta'(0)$, say $e_0 = v, e_1, e_2, e_3$.
Then $e_0$ vanishes in $M_p$.
Consequently, an arbitrary vector $\sum_{j = 0}^3 a_j e_j$ in $\tilde{M}_p$, with $a_j \in \mathbb{R}$, is mapped to $\sum_{j = 1}^3 a_j e_j$ in $M_p$. 
We therefore define $M_p$ to be the image of the degenerate metric $h$ at $p \in \tilde{M}$ (Definition \ref{internal tangent space def}):
\begin{equation*}
M_p := \im h_p \subseteq \tilde{M}_p.
\end{equation*}

We call $M_p$ a `tangent space' because it is a vector space that reduces to the tangent space $\tilde{M}_p$ of the manifold $\tilde{M}$ whenever $M = \tilde{M}$ (that is, whenever the set of pointon worldlines is empty).
Moreover, since $M_p$ is a subspace of $\tilde{M}_p$ that does not contain $v$, \textit{the (vector space) dimension of $M_p$ can be at most three:} the dimension of a vector space is the number of elements in any basis, and there is a basis for $\tilde{M}_p$ which contains $v$.

We would like to compare this tangent space dimension with the dimension of $M$ itself.
$M$ is not a manifold, however, and so we cannot apply the definition of manifold dimension given above.
Nevertheless, we define the dimension of $M$ to be the dimension of its external manifold $\tilde{M}$ since $M$ is obtained from $\tilde{M}$ by removing a locally finite set of curves in $\tilde{M}$, and such a set of curves has measure zero in $\tilde{M}$.\footnote{This is indeed the case in the setting of nonnoetherian algebraic geometry: the Krull dimension of a nonnoetherian coordinate $R$ equals the Krull dimension of its depiction $S$ since the varieties $\operatorname{Max}R$ and $\operatorname{Max}S$ are birationally equivalent \cite[Theorem 2.5]{B4}.}
In this sense, then, the dimension of $M$ is four.

Recall that the dimension of a tangent space on a manifold is always equal to the dimension of the manifold itself.
In contrast, if singularities are allowed---as in algebraic geometry---\textit{then this equality of dimensions necessarily fails}. 
In fact, a singular point of a variety is precisely a point where the tangent space dimension is larger than the dimension of the variety.\footnote{Specifically, a point $\mathfrak{m}$ of a variety $\operatorname{Max}R$ is singular if and only if the vector space dimension over $k = R_{\mathfrak{m}}/\mathfrak{m}$ of the (Zariski) cotangent space $\mathfrak{m}/\mathfrak{m}^2$ at $\mathfrak{m}$, or equivalently, the dimension of the tangent space $(\mathfrak{m}/\mathfrak{m}^2)^* = \operatorname{Hom}_k(\mathfrak{m}/\mathfrak{m}^2,k)$, is larger than the Krull dimension of the local ring $R_{\mathfrak{m}}$.}
For example, if $X$ is a plane algebraic curve, then its tangent spaces are $1$-dimensional at all smooth points of $X$ and $2$-dimensional at all singular points.
In the case of an internal spacetime $M$, we have found the opposite to hold: the tangent space dimensions $\dim M_{\beta(t)}$ along a pointon worldline $\beta \subset \tilde{M}$ are \textit{smaller} than the dimension of $M$, rather than \textit{larger}. 
Consequently, \textit{pointon worldlines are a novel type of singularity whose tangent spaces have smaller dimension, rather than larger dimension, to that of the underlying geometric space.}

\section{Ontological models} \label{HS appendix}

We briefly review the classification of ontological (hidden variable) models given by Harrigan and Spekkens \cite{HS}.
Consider a quantum system with Hilbert space $\mathcal{H}$, and suppose the system possesses an underlying ontic state space $\Lambda$.
Let $p(\lambda | P)$ be the probability distribution that an ontic state $\lambda \in \Lambda$ results from the preparation procedure $P$; and let $p(k | M, \lambda)$ be the probability distribution that the outcome $k$ results from the measurement $M$ of $\lambda$. 
Let $\rho$ be the density operator associated to $P$, and let $E_k$ be the POVM associated to the outcome $k$ of $M$.
In order for the model to reproduce quantum statistics,\footnote{For our purposes, $P$ and $M$ produce the pure states $\psi$ and $\phi$ in $\mathcal{H}$ respectively; whence $\rho = \left| \psi \right\rangle \! \left\langle \psi \right|$, $E_k = \left| \phi \right\rangle \! \left\langle \phi \right|$, and $\operatorname{tr}(\rho E_k) = \left| \left\langle \phi | \psi \right\rangle \right|^2$.}
 it must satisfy Born's rule \cite[Definition 1]{HS}:
\begin{equation*} \label{int}
p(k | M, P) := \int_{\Lambda} d \lambda \, p(k | M, \lambda) p(\lambda | P) = \operatorname{tr}(\rho E_k).
\end{equation*}

The relationship between $\Lambda$ and $\mathcal{H}$ specifies the following classes of ontological models \cite[Definitions 4 and 5]{HS}. 
\begin{itemize} 
 \item[(\textsc{a})] \textit{$\psi$-ontic:} Each ontic state $\lambda \in \Lambda$ is represented by a unique quantum state $\psi \in \mathcal{H}$.  
Thus, if $\psi \not = \phi$, then
\begin{equation} \label{psi phi}
p(\lambda | P_{\psi})p( \lambda | P_{\phi} ) = 0.
\end{equation}
There are two subclasses of $\psi$-ontic models:
\begin{itemize}
 \item[(i)] \textit{$\psi$-complete:} Each quantum state $\psi$ represents a unique ontic state $\lambda$.
 \item[(ii)] \textit{$\psi$-supplemented:} There is a quantum state $\psi$ that represents more than one ontic state in $\Lambda$.
\end{itemize}
 \item[(\textsc{b})] \textit{$\psi$-epistemic:} There is an ontic state $\lambda$ that is represented by more than one quantum state in $\mathcal{H}$; thus (\ref{psi phi}) does not hold for $\lambda$.
\end{itemize}

A quantum state $\psi$ is therefore $\psi$-ontic if and only if a variation of $\psi$ implies a variation of reality, and $\psi$-epistemic if and only if a variation of $\psi$ does not imply a variation of reality.
In terms of mappings, a model is $\psi$-complete if the correspondence $\Lambda \to \mathbb{P}\mathcal{H}$ is bijective; $\psi$-supplemented if the correspondence $\Lambda \to \mathbb{P}\mathcal{H}$ is surjective but not injective; and $\psi$-epistemic if the opposite correspondence $\mathbb{P}\mathcal{H} \to \Lambda$ is surjective but not injective.

The Kochen-Specker model (see Section \ref{tangent space projections}) is a $\psi$-epistemic model since the product
\begin{equation*}
p(\bm{\lambda} | \bm{\psi}) p(\bm{\lambda} | \bm{\phi}) = \tfrac{1}{\pi^2} H(\bm{\psi} \! \cdot \! \bm{\lambda}) H(\bm{\phi} \! \cdot \! \bm{\lambda} ) (\bm{\psi} \! \cdot \! \bm{\lambda})( \bm{\phi} \! \cdot \! \bm{\lambda})
\end{equation*}
is nonzero for some $\bm{\lambda}$ whenever $\bm{\psi}$ and $\bm{\phi}$ are not orthogonal.

\ \\
\textbf{Acknowledgments.}
The author thanks an anonymous referee for their careful reading and helpful comments.
The author was supported by the Austrian Science Fund (FWF) grant P 34854.

\bibliographystyle{hep}

\end{document}